\documentstyle[11pt,aspconf]{article}
\begin{document}

\title{Shocking Outflows in Seyfert Galaxies:\\
       An Imaging Perspective}
\author{Richard W.\, Pogge}
\affil{Department of Astronomy, The Ohio State University,
       174 W. 18th Ave, Columbus, OH 43210-1106, USA}


\begin{abstract}
Representative results of a search for structures suggestive of shocked gas
in the circumnuclear regions of Seyfert galaxies are presented.  At issue
is whether we can identify regions that appear to be shocked by outflows
from the active nucleus.  Such regions will be good targets for future
spectrophotometric studies to determine what r\^ole shocks play in the
physics of the AGN/host galaxy interaction.
\end{abstract}



\section{Introduction}

There has been much debate at this colloquium over the importance of shocks
in AGNs.  Specifically, is the physics (ionization, heating, etc.) of the
narrow-line region and the spatially extended emission-line regions
dominated by shocks or merely aided and abetted by them?  Shocks are often
invoked to help explain various anomalies relative to the predictions of
pure photoionization models that are observed in the temperature,
kinematics, and ionization state of the extended emission-line gas,
although they are not the only explanations available.  Further, one
expects shocks to play {\it some} r\^ole, since shocks are practically
ubiquitous in outflow sources seen at more modest scales in our own Galaxy,
admittedly with varying degrees of energetic importance.  An excellent
review of the basic issues may be found in Morse et al.\,(1996), as well as
papers in these proceedings (esp. by Viegas, Wilson, Evans, Allen, and
Bicknell).

I want to step back a little from the theoretical issues and ask the
question: {\it Is there any morphological evidence of shocked gas in
Seyfert galaxies?}  In particular, does imaging reveal any gas structures
within the circumnuclear regions of AGNs where we might suspect that shocks
play a r\^ole, whether as the principal ionization source, or as an agent
for organizing gas that is otherwise primarily photoionized by the nucleus
into the shapes that we see?

\section{Search Strategy}

In outline, the strategy is to start with the extensive ground-based
emission-line imaging and spectroscopic surveys of Seyfert nuclei (e.g.,
Pogge 1989; Mulchaey et al.\,1996; Whittle 1992a,b), and select objects
based on suggestive emission-line morphologies or evidence of disturbed
kinematics.  By ``suggestive morphologies'' I mean ionized gas regions that
resemble in form (if not in scale) the shocked regions well-known from
Galactic stellar bipolar outflow sources (e.g., bowshocks and jet shocks as
seen in Herbig-Haro objects) and supernova remnants.  Jet shocks might be
expected if the radio jets seen in some Seyferts entrain surrounding
material.  Bowshocks are expected to form where nuclear outflows ram into
the surrounding ISM, and cooling-length arguments suggest that they should
be geometrically thin and edge-brightened, with their thicknesses
unresolved even at {\it HST} resolutions.  As such, relatively ``sharp''
features in ground-based images should become even sharper in {\it HST}
images, and emerge in higher contrast against the diffuse starlight or
emission-line gas backgrounds in which we expect them to be embedded.  The
second step, then, has been to collect archival {\it HST} images of
galaxies selected out of the ground-based surveys, giving preference to
FOC/COSTAR or WFPC2 images (avoiding the difficulties of image
reconstruction).

Such a ``sample'' is of course fairly biased, and does not constitute a
survey for shocked structures {\it per se}.  Indeed, it is important to
emphasize that in general the collection of Seyferts imaged thus far by
{\it HST} is strongly biased towards those galaxies previously suspected of
being ``interesting'' by various investigators for a variety of reasons,
and thus no statistical conclusions may be drawn from these data in any
event.

A surprise from the archival {\it HST} imaging data examined thus far has
been how useful the {\it broad-band} images are for this search, contrary
to my prejudice that narrow-band filter images are best.  In a number of
cases, sharply defined, high surface brightness emission regions appear in
relatively high contrast against the stellar background of the host
galaxies in broad ($>1000$\AA) filter images (see Fig.~2).  A further
important benefit of these images is that any dust structures or spiral
arms continuing to curl in towards the nucleus proper are readily visible,
whereas such features are usually all but lost in narrowband images.  This
greatly expands the set of galaxies we may consider, as there is much more
broad-band imaging available compared to narrowband imaging (in consequence
of the fact that the restricted set of narrowband filters available on {\it
HST} limits which objects may be studied in this way).

\section{Representative Results}

I show here three examples representative of the types of shock-like
morphologies found among the Seyferts studied.  These are the best cases
found thus far.

\noindent{\bf Markarian 3}: This galaxy, studied in detail by Capetti et 
al.\,(1995) shows the best example seen thus far of a jet shock.  The
narrow radio jet (Kukula et al.\,1993) lies precisely along the axis of the
``bar'' of emission-line clouds in the inner 1\arcsec\ (Fig 1).  No
electronic version of the radio map was available for plotting here, but
see Capetti et al.\ for a contour map overlay.

\noindent{\bf Markarian 573}: The relation between the 6cm radio continuum
structure (Ulvestad \& Wilson 1984) and the bowshock-like emission-line
structure has been shown by Pogge \& DeRobertis (1995).  In Fig.~2 (left),
a WFPC2 image in a filter including [O~III] is shown after subtraction of a
broad I-band (F814W) image.  This ``continuum'' image over-subtracts
starlight in regions of dust extinction, resulting in ``negative'' features
at dust lanes which map to white colors.  The radio nucleus is located (by
physically well-motivated fiat, not astrometry) at the center of the
nuclear spiral dust lanes, not the [O~III] peak.  This is probably the best
example of an optical emission-line bowshock in a Seyfert (the best radio
bowshock is in NGC\,1068; Wilson \& Ulvestad 1987).

\begin{figure}
 \plotone{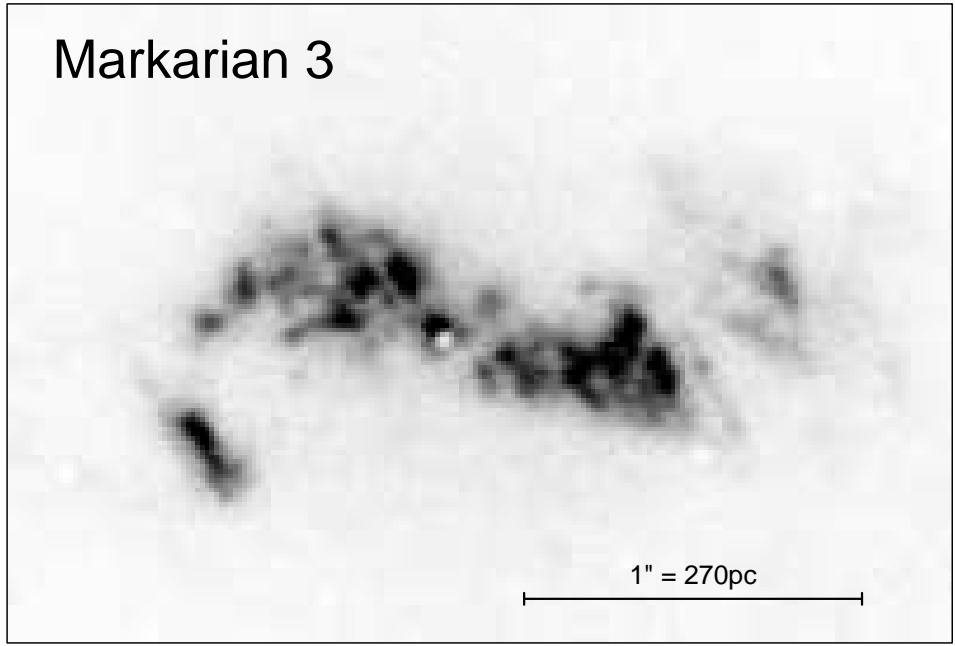}
 \caption{A representative jet shock morphology: FOC (COSTAR) [O\,III]
 image of Mrk\,3.  The MERLIN radio jet map follows the central bar of
 emission-line clouds (not shown).  The nucleus is partly blocked by a
 Riseau mark.  See Capetti et al.\,(1995) for a detailed discussion.}
\end{figure}

\begin{figure}
 \plotone{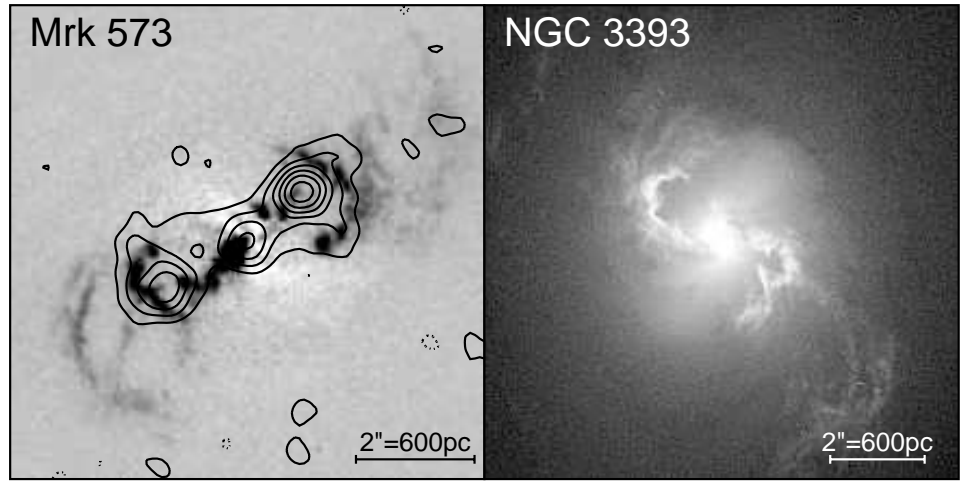}
 \caption{Bowshock-like morphologies: {\it Left:} Mrk\,573: continuum
  subtracted WFPC2 [O\,III] image with VLA 6cm radio contours superimposed.
  Emission clouds are dark and dust features appear white.  {\it Right:}
  NGC\,3393: WFPC2 F606W image, shown with emission white to improve
  contrast.  The scales are indicated in each panel.}
\end{figure}

\noindent{\bf NGC\,3393}: Figure 2 (right) shows a WFPC2 F606W image
of the central 10\arcsec.  The nucleus is at the apex of dusty nuclear
spiral arms, and the S-shaped emission-line structure follows these same
arms.  The bowshock-like arcs are reminiscent of those in Mrk\,573 (left)
and of similar scale, but their relation to the spiral dust lanes makes it
as likely that we are seeing dense arm gas and dust illuminated by ionizing
photons emerging from the active nucleus.

\section{Summary}

The imaging presented here shows representative examples of structures that
satisfy our basic notions of what regions of shocked gas should look like.
We see structures suggestive of both jets of material in the outflow
itself, and bowshocks where it appears that nuclear outflows are impinging
on the surrounding interstellar medium, although NGC\,3393 has suspiciously
located dust features that make interpretation of its arcs as bowshocks
less secure. There is, however, forthcoming unpublished work (Baldwin;
Lawrence priv comm) that suggests it may really be a close cousin of
Mrk\,573, so it is yet not clear how much the local environment can serve
to confound our morphological judgment.

Are there shocks in Seyferts?  I believe the answer is ``yes''.  Do these
shocks do more than sculpt the circumnuclear gas?  This is a question that
only detailed spectrophotometry to look for tell-tale ionization and
kinematic signatures will address.  Whether these spectra will provide
definitive answers remains to be seen, a conclusion which should come as a
shock to nobody.

\acknowledgments

Support for this work was provided by NASA through grant number
AR-06380.01-95A from the Space Telescope Science Institute, which is
operated by the Association of Universities for Research in Astronomy,
Inc., under NASA contract NAS5-26555.

\end{document}